\documentstyle[prd,aps,amssymb,epsf,preprint,tighten]{revtex}
\draft
\begin{document}

\title{
\rightline{\small{\em Phys. Rev. D {\bf 64}, 024007 (2001)}}
Gravitational waveforms with controlled accuracy}

\author{Roberto G\'{o}mez\thanks{E-mail address: gomez@pitt.edu}}

\address{Department of Physics and Astronomy,
         University of Pittsburgh, Pittsburgh, Pennsylvania 15260}

\maketitle

\begin{abstract}

A partially first-order form of the characteristic formulation is
introduced to control the accuracy in the computation of gravitational
waveforms produced by highly distorted single black hole spacetimes.  Our
approach is to reduce the system of equations to first-order differential
form on the angular derivatives, while retaining the proven radial and
time integration schemes of the standard characteristic formulation.
This results in significantly improved accuracy over the standard
mixed-order approach in the extremely nonlinear post-merger regime of
binary black hole collisions.

\end{abstract}

\pacs{04.20.Ex, 04.25.Dm, 04.25.Nx, 04.70.Bw}

\section{Introduction}

One of the most impressive achievements of the characteristic
approach is its demonstrated ability to carry out long-term stable
evolutions~\cite{stable} of generic single black hole spacetimes. Work is
under way to extend these numerical simulations to the post-merger regime
of binary black hole coalescence, by first computing the gravitational
radiation emitted during a white hole fission~\cite{white}, and then
extending these results to compute the gravitational waveforms
emitted during the post-merger phase of a binary black hole
collision~\cite{merger}.

The present work is an outgrowth of one of these projects, having
been motivated by numerical difficulties that we have encountered with
the present implementation of the characteristic code~\cite{news}. In
the course of our simulations of a white-hole fission~\cite{white},
we have noticed an angular oscillation mode in some of the metric
variables of the characteristic formulation, similar in appearance
to a {\it red-black}~\cite{recipes} decoupling. Numerical experiments
reveal that although the area that it affects can be somewhat reduced,
its amplitude does not converge to zero with increasing grid resolution
for practical grid sizes in the rapidly changing environment near the 
time of merger of a binary black hole.
We had expected to see convergence in this regime, given the measured 
second order convergence of the code on less rapidly changing solutions, 
even at fairly low resolutions~\cite{news,matter}.

This effect is clearly of numerical nature, and would not have been
observed in earlier work, as it is only triggered when, in addition to
extreme nonlinearities in the metric functions (such as those found in
the post-merger phase of the head-on collision of two black holes, or
in the time-reversed picture, during the fission of a single
black-hole), a fairly large degree of asymmetry and of time dependence
in the metric functions is present. Moreover, it appears that even when
some of these conditions are satisfied, the dissipation built into the
algorithm used for the evolution the conformal metric of the
spheres~\cite{disip} can, in certain circumstances, be sufficient to
suppress the spurious oscillations, provided the boundary data is
specified analytically~\cite{moving}.

In the systems that we have previously considered, such as
Robinson-Trautman spacetimes, a perturbed Kerr black hole~\cite{moving}
and the collapse of matter into a black hole~\cite{matter}, our choice
of the gauge conditions at the boundary effectively precluded fulfilling
the second condition above. In all cases, the boundary data was given
analytically, so no extraneous effects could be observed.

This oscillatory mode is not an {\it unstable} mode of the finite
difference approximation in the usual sense of exponential growth of
the error. This is confirmed by tests where random initial and/or
boundary data~\cite{cce} is prescribed, and the code run for many
crossing times. But it is worth pointing out that while those tests
are useful to spot exponentially growing modes, they would not detect
a non-exponentially growing oscillatory mode such as the one we address
here.

We observe, in fact, an inaccuracy in the computation of the metric
variable $W$, which does not converge to zero, and gets triggered in
situations where there are large angular variations and rapid time
changes in the metric variables, specially in $\beta$, Eq.~(\ref{eq:bmet}). 
Because $\beta$ is related to the time slicing of the boundary 
(and in the black hole case, this boundary is the event horizon ${\cal 
H}^{+}$), $\beta$ will indeed change substantially 
near the time of {\it coalescence} of a binary black hole horizon (or, 
in the time-reversed description actually used by our numerical code, 
near the time of {\it fission} of a white hole).

The PITT null code runs stably in the usual sense (without exponentially
growing modes) in the configurations that we have explored. The effect
of the angular noise is nevertheless important for our present purposes,
since the angular error it introduces in the metric functions is most
pronounced near null infinity, and this in turn spoils the accuracy of
the waveforms computed.

We have found that the difficulty lies with the characteristic evolution
equations themselves, and it is not a boundary-induced effect. Although
the boundary (horizon) data is smooth and, by construction, satisfies
the consistency conditions~\cite{horizon}, the evolution and hypersurface
equations themselves introduce the above mentioned oscillation mode, which
is shown clearly in Fig.~\ref{fig:Wold}. We have found clear evidence
that the origin of the difficulty lies with those terms in the equations
that contain second angular derivatives.  These type of terms are present
both in the evolution equation for the conformal metric of the surfaces at
constant luminosity distance $r$, and in the hypersurface equation for the
metric function $W$ itself. Most notably, when this terms are suppressed,
so is the angular oscillation mode in $W$.  The most troublesome terms
are those containing second angular derivatives of the metric variable
$\beta$, of the form $\eth\bar\eth \beta$ and $\eth^2\beta$.

Second angular derivatives of the conformal metric $h_{AB}$, 
Eq.~(\ref{eq:bmet}), also appear
in the hypersurface equation for $W$, in the form of ${\bar\eth}^2$ and
$\eth\bar\eth$ operators acting on quantities defined in terms of the
conformal metric. However, these terms do not seem to lead to numerical
difficulties as pronounced as those caused by the terms involving second
angular derivatives of $\beta$, although the reason behind this is not
entirely clear at the moment. It is possible that in the present regime
the contribution from these terms is not as important.

Our rationale for implementing the characteristic code in mixed first
and second order form has been that doing so is entirely straightforward,
and leads to an accurate and stable discretization with the least number
of variables. The argument of using the least variables is compelling;
however one must consider whether relaxing this requirement might not
lead to better overall numerical behavior. We will show that this is
indeed the case.

Rigorous stability arguments can be given for the linearized
system~\cite{axisymmetric} in the axisymmetric case, with the analysis
extending unmodified to the full three-dimensional problem as shown
in~\cite{cce}. Later work~\cite{disip} has shown that special care must
be taken in the nonlinear case, as some unstable modes are present that
are not revealed by the Von-Neuman analysis of the linear problem. One
must keep in mind however that a stable discretization of a model
problem as in~\cite{disip} may not address other issues such as those
raised here. The numerical analysis for mixed order systems is not yet
at the level of sophistication achieved for first differential order
systems. Long-term stable evolutions~\cite{stable} have been achieved,
and this is a consequence of the remarkably good choice of variables
and coordinates inherent to Bondi's system of equations, as well as of
the care taken in implementing them numerically.

We have attempted to solve the problem at hand by the expeditious method
of introducing further dissipation in the equations, in the spirit
of~\cite{disip}. Regrettably, these numerical remedies do not seem to
have the desired effect, apparently because they do not go to the root
of the problem.

We consider here a different solution, which is analytic in nature, but
motivated by a numerical application, specifically the need to attain
controlled accuracy in the computation of gravitational waveforms out of
highly distorted single black hole systems. Our approach is to remove
the offending second angular derivatives from our system of equations
altogether, by introducing additional variables, in terms of which the
enlarged system can be written in first differential order form in the
angular coordinates. We thus implement a {\it partial} reduction to a
first differential order system, leaving the radial part of the evolution
equations in second order form, and allowing for mixed radial and angular
derivatives, both in the original equations and in the auxiliary equations
introduced. One goal is to preserve, where appropriate, the radial and
time integration schemes developed for the PITT null code, whose long-term
stability (in the usual sense of not having exponentially growing modes)
has been amply demonstrated~\cite{stable}. At the same time, we perform
the necessary modification to the system of equations to make the PITT
null code useful for the problem at hand.

The plan of the present paper is as follows: in Sec.~\ref{sec:2nd}
we briefly review the characteristic formulation in its standard
second-order form. This material has appeared in Refs.~\cite{news,cce},
and we include it here in a condensed form, as needed for the derivation
of the partially reduced system. Our new results are contained in
Sec.~\ref{sec:reduction} and onward, where we present a partial
reduction of the characteristic formulation, with all second order
angular derivatives eliminated with the aid of a (minimal) set of
additional variables. In Sec.~\ref{sec:numeric}, we review the
discretization strategy for the system of equations introduced in
Sec.~\ref{sec:reduction}. Finally, we present an application of
the partially reduced system in Sec.~\ref{sec:applications} to the
computation of the spacetime exterior to a white hole horizon, comparing
our results with those obtained with the standard null cone formulation.

\section{Null cone formulation in second order form}
\label{sec:2nd}

The material in this section has appeared in Refs.~\cite{news,cce} which
should be consulted for a detailed derivation. Here we present the main
equations (including all the nonlinear terms), in a slightly more compact
form than that of Ref.~\cite{news}, for ease of reference and to provide a
starting point for the approach developed in Sec.~\ref{sec:reduction}.

We use coordinates based upon a family of outgoing null hypersurfaces,
letting $u$ label these hypersurfaces, $x^A$ $(A=2,3)$, label the null
rays, and $r$ be a surface area coordinate, such that in the
$x^\alpha=(u,r,x^A)$ coordinates, the metric takes the Bondi-Sachs
form~\cite{bondi,sachs}

\begin{eqnarray}
   ds^2 & = & -\left[e^{2\beta}\left(1 + {W \over r}\right)
              - r^2h_{AB}U^AU^B\right]du^2
              - 2e^{2\beta}dudr
 \nonumber \\
            & - & 2r^2 h_{AB}U^Bdudx^A
              + r^2h_{AB}dx^Adx^B,    \label{eq:bmet}
\end{eqnarray}
where $W$ is related to the Bondi-Sachs variable $V$ by $V=r+W$; and
where $h^{AB}h_{BC}=\delta^A_C$ and $det(h_{AB})=det(q_{AB})$, with
$q_{AB}$ a unit sphere metric, given in terms of a complex dyad $q_A$
satisfying $q^Aq_A=0$, $q^A\bar q_A=2$, $q^A=q^{AB}q_B$, with
$q^{AB}q_{BC}=\delta^A_C$ and $ q_{AB} =\frac{1}{2}\left(q_A \bar
q_B+\bar q_Aq_B\right)$.

We represent tensors on the sphere by spin-weighted
variables~\cite{eth}. The conformal metric $h_{AB}$, which satisfies
the condition $\det(h_{AB})=\det(q_{AB})$, is represented by the
complex function $J=\frac{1}{2}h_{AB}q^Aq^B$, and by the real function
$K=\frac{1}{2}h_{AB}q^A\bar q^B$, where $K^2=1+J\bar J$. The metric
functions $U^A$ are similarly encoded in the complex function $U=q_A
U^A$~\cite{news,cce}. Angular derivatives are expressed in terms of
$\eth$ and $\bar\eth$ operators, for details, see Ref.~\cite{eth}.

The equations for the null cone formulation, in second order form, are
obtained directly from the appropriate components of the Ricci
tensor~\cite{news,cce}. These equations form a hierarchy, splitting
into hypersurface equations, which involve only derivatives on the null
cone, {\it i.e.} $R_{rr}=0$ provides an equation for $\beta_{,r}$ in
terms of $J$, while $R_{rA}q^A=0$ gives $U_{rr}$ in terms of $J$ and
$\beta$, and the trace $R_{AB}h^{AB}=0$ yields $W_{,r}$ in terms of
$J$, $\beta$ and $U$, respectively. 

The main equations~\cite{news,cce} consist of the hypersurface equations

\begin{eqnarray}
      \beta_{,r} &=& \frac{r}{8}\left(J_{,r}\bar J_{,r}-K^2_{,r} \right),
     \label{eq:beta_r} \\
     (r^2 Q)_{,r}  &=& -r^2 (\bar \eth J + \eth K)_{,r}
                +2r^4\eth \left(r^{-2}\beta\right)_{,r}
     \nonumber \\
&+& r^2 \Bigg( (1-K) ( \eth K_{,r} + \bar \eth J_{,r} ) + \eth (\bar J
J_{,r} ) + \bar \eth ( J K_{,r} )  - J_{,r} \bar \eth K \nonumber \\
   & + & \frac{1}{2K^2}(\eth \bar J (J_{,r} - J^2 \bar J_{,r} ) + \eth J (\bar
J_{,r} -\bar J^2  J_{,r} ) ) \Bigg).
     \label{eq:Q_r} \\
   U_{,r}  &=& \frac{e^{2\beta}}{r^2} \left(KQ-J\bar Q \right),
     \label{eq:U_r} \\
   W_{,r} &=& \frac{1}{2} e^{2\beta}{\cal R} -1
- e^{\beta} \eth \bar \eth e^{\beta}
+ \frac{1}{4} r^{-2} \left(r^4
                           \left(\eth \bar U +\bar \eth U \right)
                     \right)_{,r} \nonumber \\
   &+& e^{2 \beta} \Bigg[ (1-K) ( \eth \bar \eth \beta
                                +  \eth \beta \bar \eth \beta)
      + \frac{1}{2} \bigg ( J (\bar \eth \beta)^2
                           + \bar J (\eth \beta)^2
                    \bigg)
\nonumber \\
& - & \frac{1}{2} \bigg ( \eth \beta ( \bar \eth K - \eth \bar J)
                          + \bar \eth \beta ( \eth K - \bar \eth J )
                  \bigg)
    + \frac{1}{2} \bigg ( J \bar \eth^2 \beta
+ \bar J \eth^2 \beta  \bigg)
     \Bigg ]
\nonumber \\
& & - e^{-2 \beta} \frac{r^4}{8} ( 2 K U_{,r} \bar U_{,r} + J \bar U^2_{,r} +
\bar J U^2_{,r}).
\label{eq:W_r}
\end{eqnarray}
and the evolution equation for $J$, obtained from $R_{AB}q^Aq^B$,

\begin{eqnarray}
    && 2 \left(rJ\right)_{,ur}
    - \left(r^{-1}V\left(rJ\right)_{,r}\right)_{,r} = \nonumber \\
    && -r^{-1} \left(r^2\eth U\right)_{,r}
    + 2 r^{-1} e^{2\beta} \left(\eth^2 \beta + (\eth\beta)^2 \right)
    - \left(r^{-1} W \right)_{,r} J
    + N_J.
    \label{eq:wev}
\end{eqnarray}
where~\cite{eth}
\begin{equation}
{\cal R} =2 K - \eth \bar \eth K + \frac{1}{2}(\bar \eth^2 J + \eth^2 \bar J)
          +\frac{1}{4K}(\bar \eth \bar J \eth J - \bar \eth J \eth \bar J).
     \label{eq:calR}
\end{equation}
and we have used the intermediate variable $Q=r^2e^{-2\beta}\left(J\bar
U_{,r}+ K U_{,r}\right)$. $N_J$ denotes the non-linear terms

\begin{equation}
N_J=N_{J1}+N_{J2}+N_{J3}+N_{J4}+N_{J5}+N_{J6}+N_{J7}+\frac{J}{r}(P_1+P_2
+P_3+P_4)
\label{eq:J_u}
\end{equation}
where
\begin{eqnarray}
N_{J1}&=& - \frac{e^{2 \beta}}{r} \bigg ( K ( \eth J \bar \eth \beta + 2 \eth K
\eth \beta - \bar \eth J \eth \beta) + J ( \bar \eth J \bar \eth \beta - 2 \eth
K \bar \eth \beta) - \bar J \eth J \eth \beta \bigg) ,
\nonumber \\
N_{J2}&=& -\frac{1}{2} \bigg ( \eth J ( r \bar U_{,r} + 2 \bar U) + \bar \eth J
( r U_{,r} + 2 U) \bigg) ,
\nonumber \\
N_{J3}&=& (1-K) ( r  \eth U_{,r} + 2 \eth U) - J ( r \eth \bar U_{,r} + 2 \eth
\bar U) ,
\nonumber \\
N_{J4}&=& \frac{r^3}{2} e^{-2 \beta} \bigg( K^2 U^2_{,r} + 2 J K U_{,r} \bar
U_{,r} + J^2 \bar U^2_{,r} \bigg) ,
\nonumber \\
N_{J5}&=& - \frac{r}{2} J_{,r} ( \eth \bar U + \bar \eth U) ,
\nonumber \\
N_{J6}&=& r \Bigg( \frac{1}{2} ( \bar U \eth J + U \bar \eth J ) (J \bar J_{,r}
- \bar J J_{,r} )  \nonumber \\
    & & + ( J K_{,r} - K J_{,r} ) \bar U \bar \eth J
     - \bar U ( \eth J_{,r} - 2 K \eth K J_{,r} + 2 J \eth K K_{,r} ) \nonumber
\\
    & & - U ( \bar \eth J_{,r} - K \eth \bar J J_{,r} + J \eth \bar J  K_{,r} )
\Bigg) ,
\nonumber \\
N_{J7}&=& r ( J_{,r} K -  J K_{,r} ) \bigg ( \bar U ( \bar \eth J - \eth K ) +
U ( \bar \eth K - \eth \bar J ) \nonumber \\
       & & + K ( \bar \eth U - \eth \bar U ) + ( J \bar \eth \bar U - \bar J
\eth U ) \bigg) ,
\nonumber \\
P_1 &=& r^2 \bigg ( \frac{J_{,u}}{K} (\bar J_{,r} K - \bar J K_{,r} ) +
\frac{\bar J_{,u}}{K} ( J_{,r} K -  J K_{,r} ) \bigg)
	    - 8 V \beta_{,r} ,
\nonumber \\
P_2 &=& e^{2 \beta} \Bigg( - 2 K ( \eth \bar \eth \beta + \bar \eth \beta \eth
\beta) - ( \bar \eth \beta \eth K
                     + \eth \beta  \bar \eth K) \nonumber \\
                & + &   \bigg( J ( \bar \eth^2 \beta + (\bar \eth \beta)^2 ) +
                           \bar J ( \eth^2 \beta + (\eth \beta)^2 ) \bigg)
               + ( \bar \eth J \bar \eth \beta + \eth \bar J \eth \beta) \Bigg),
\nonumber \\
P_3 &=& \frac{r}{2} \bigg( ( r \bar \eth U_{,r} + 2 \bar \eth U) +
		       ( r  \eth \bar U_{,r} + 2 \eth \bar U) \bigg) ,
\nonumber \\
P_4 &=& - \frac{r^4}{4} e^{-2 \beta} ( 2 K U_{,r} \bar U_{,r} + J \bar U^2_{,r}
		      + \bar J U^2_{,r} ).
\label{eq:nji}
\end{eqnarray}

In Refs.~\cite{news,cce} we have made a partial reduction to a first
order system by introducing the auxiliary variable $Q$, which plays the
role of $U_{,r}$. We sometimes refer to this system as
{\it first order} (in time) because only the first (retarded) time
derivative of $J$ appears. See Refs.~\cite{news,cce} for details of the
discretization of the above system of equations.

\section{Null cone formulation in partially reduced form}
\label{sec:reduction}

A reduction to first order form of the system of equations in
Ref.~\cite{cce}, in the quasi-spherical approximation, was considered by
Frittelli and Lehner~\cite{sfll}. In their work, the main variables are
$J_{AB}=h_{AB}-q_{AB}$, $U_{A}$, $\beta$ and ${\hat W}=V-r+2m$, which measure
deviation from Schwarzschild. The auxiliary variables they introduce
to reduce the system to first order form are the following radial
and angular derivatives of the metric functions, $B_{A}=\beta_{,A}$,
$M_{ABC}=J_{AB,C}$, $Q^{A}=U^{A}_{,r}$ and $P_{AB}=J_{AB,r}$.

Here we do not work with tensor components directly, but instead,
following Refs.~\cite{cce,news}, we represent tensors in terms of
spin-weighted variables. For example, instead of writing the equations
in terms of $B_{A}$, we work with $B=q^{A} \beta_{,A}=\eth\beta$,
and express its angular derivatives in terms of $\eth$ and $\bar\eth$
operators. Second angular derivatives of $\beta$ can in turn be expressed
in terms of $\eth B$ and $\bar\eth B$.

Eliminating all second angular derivatives in the hypersurface and
evolution equations necessitates the introduction of exactly three complex
variables. Since by symmetry there are only three independent components
of the conformal metric $h_{AB}$, there are then only six possible angular
derivatives $h_{AB,C}$, which are encoded in the three complex quantities

\begin{eqnarray}
 \mu & =     \eth J & = \frac{1}{2}h_{AB,C} q^A q^B q^C, \\
 \label{eq:mudef}
 \nu & = \bar\eth J & = \frac{1}{2}h_{AB,C} q^A q^B \bar q^C, \\
 \label{eq:nudef}
  k  & = \eth K & = \frac{1}{2}h_{AB,C} q^A \bar q^B q^C,
 \label{eq:kdef}
\end{eqnarray}
with spin weight $(3)$, $(1)$ and $(1)$, respectively. The first
order system for the quasi-spherical case considered in~\cite{sfll}
introduces six new variables just for the angular derivatives of the
conformal metric because no use is made there of the determinant
condition. The conformal metric $h_{AB}$ satisfies the condition
$\det(h_{AB})=\det(q_{AB})$, {\it i.e.} it contains only two independent
pieces of information, which are given in our notation by the complex
function $J=\frac{1}{2}h_{AB}q^Aq^B$. The remaining dyad component, given
by the real function $K=\frac{1}{2}h_{AB}q^A\bar q^B$, is completely
determined by the relation $K^2=1+J\bar J$, which is a consequence of
the determinant condition.

Therefore only two of the three complex quantities in
Eqs.~(\ref{eq:mudef}),(\ref{eq:kdef}) can be given independently, with
the remaining one fixed by the determinant condition.  Our choice is to
introduce the additional variables $\nu=\bar\eth J$ and $k=\eth K$. By
inspection of Eqs.~(\ref{eq:beta_r})--(\ref{eq:J_u}), we note that to put
the system in first order form in the angular derivatives, we need only
introduce one more auxiliary variable, $B=\eth\beta$.

The new variables are initialized at the boundary with $B=\eth\beta$,
$\nu=\bar\eth J$ and $k=\eth K$ respectively, and the consistency
conditions

\begin{eqnarray}
     B_{,r} & = & \eth\beta_{,r} , \\
   \nu_{,r} & = & \bar\eth J_{,r} , \\
     k_{,r} & = & \eth K_{,r}
\label{eq:Bnuk}
\end{eqnarray}
propagate them to the interior, being the additional hypersurface
equations for the new variables.

The complete system of equations consists of Eqs.~(7)--(12) of
Ref.~\cite{news}, together with Eq.~(\ref{eq:Bnuk}), {\it i.e.},
the evolution equation for $J$

\begin{eqnarray}
     2 \left(rJ\right)_{,ur}
    - \left(r^{-1}V\left(rJ\right)_{,r}\right)_{,r} &=&
     - K \left( r \eth U_{,r} + 2\, \eth U \right)
    + \frac{2}{r} e^{2\beta} \left( \eth B + B^2 \right)
\nonumber \\
    & - & \left(r \tilde{W}_{,r} + \tilde{W} \right) J
      + J_{H} + J P_u ,
    \label{eq:J}
\end{eqnarray}
and the hierarchy of hypersurface equations

\begin{eqnarray}
   \nu_{,r} &=& \bar\eth J_{,r}
   \label{eq:nu} \\
   k_{,r} &=& \eth K_{,r}
   \label{eq:k} \\
    \beta_{,r} &=& \frac{r}{8} \left( J_{,r} \bar J_{,r} - K^2_{,r}\right)
   \label{eq:beta} \\
   B_{,r} &=& \eth\beta_{,r}
   \label{eq:B} \\
     (r^2 Q)_{,r}  &=&
      r^2 \bigg[ - K ( k_{,r} + \nu_{,r} )
                 + \bar \nu J_{,r} + \bar J \eth J_{,r}
                 + \nu K_{,r} + J \bar k_{,r}
- J_{,r} \bar k
           \bigg]  \nonumber \\
   & + &  \frac{r^2}{2K^2}
          \left[ \bar \nu \left(J_{,r} - J^2 \bar J_{,r} \right)
                + \eth J \overline{\left(J_{,r} - J^2 \bar J_{,r} \right)}
          \right] +  2 r^2 B_{,r} -4r B
     \label{eq:Q} \\
   r^2 U_{,r}  &=& e^{2\beta} \left( K Q - J \bar Q \right)
     \label{eq:U} \\
     (r^2 \tilde{W})_{,r}  &=& \Re \Bigg\{ e^{2\beta}
    \left( \frac{\cal R}{2} -K \left( \bar\eth B + B \bar B\right)
   + \bar J \left( \eth B + B^2 \right) + (\nu - k) \bar B \right)
\nonumber \\
 & - & 1 + 2\,r \bar \eth U + \frac{r^2}{2} \bar\eth U_{,r}
   -  e^{-2 \beta} \frac{r^4}{4}
 \bar U_{,r} \left( K U_{,r} + J \bar U_{,r} \right) \Bigg\}
 \label{eq:W}
\end{eqnarray}
where
\begin{equation}
  {\cal R} = \Re \left( 2 K + \bar \eth \left(\nu - k \right)
  + \frac{1}{4K} \left( |\eth J|^2 - |\nu|^2 \right) \right) .
   \label{eq:Ricci}
\end{equation}
In Eqs.~(\ref{eq:J})--(\ref{eq:Ricci}) we have expressed the second angular
derivatives in terms of $\eth$ operators acting on $B$, $\nu$ and $k$,
replacing also $\eth$ and $\bar\eth$ operators acting upon $J$, $K$ and
$\beta$ with the corresponding auxiliary variables whenever possible. 

As in~\cite{cce}, we have regularized the $W_{,r}$ equation by setting
$W=r^2\tilde{W}$, and as in Appendix A of~\cite{news}, we split the
terms which vanish in the quasi-spherical approximation~\cite{cce}
into two parts, one which contains only hypersurface derivatives,

\begin{eqnarray}
J_{H}&=& \frac{e^{2 \beta}}{r}
   \Bigg( - K \eth J \bar B + \left( K \nu + (K^2-1) \eth J - 2 K k \right) B 
          \nonumber \\
  &+& J \Big[ ( 2k - \nu) \bar B - 2 K (\bar\eth B + B \bar B)
   + 2 \Re \left[ (\nu - k)\bar B + \bar J \left(\eth B + B^2 \right) \right] 
        \Big]
   \Bigg)
      \nonumber \\
&+& \frac{r^3}{2} e^{-2 \beta} \Bigg(
          \left( K U_{,r} + J \bar U_{,r} \right)^2
         - J \Re \Big[ \bar U_{,r} \left( K U_{,r} + J \bar U_{,r} \right) 
                 \Big]
     \Bigg)
      \nonumber \\
  &-&\frac{1}{2} \Big[  \nu \left( r      U_{,r} + 2      U \right) 
                      + \eth J \overline{\left( r U_{,r} + 2 U \right)} 
                 \Big]
  + J i \Im [\bar\eth \left(r U_{,r} + 2 U \right)]
  - rJ_{,r} \Re [\bar\eth U] 
  \nonumber \\
  &+& r \left(\bar U \eth J + U \nu \right) i \Im \left[ J \bar J_{,r} \right]
   - r \left( \bar U \eth J_{,r} + U \nu_{,r} \right)
  \nonumber \\
  &-& 2 r \left( J K_{,r} - K J_{,r} \right) 
      \left( \Re[\bar U k] + i \Im[K \bar\eth U - \bar J \eth U]
      \right)
\nonumber \\
      &-& 8 J \left(1 + r \tilde{W} \right) \beta_{,r} \ ,
\label{eq:J_H}
\end{eqnarray}
and another, where we have explicitly isolated the only non-linear term 
where (retarded) time derivatives of $J$ appear,

\begin{equation}
P_u = \frac{2\,r}{K}
\Re\left[ J_{,u} \left( \bar J_{,r} K - \bar J K_{,r}\right) \right] .
\label{eq:p_u}
\end{equation}

The procedure just described eliminates all second angular derivatives
from the hypersurface and evolution equations. In computing the
Bondi News, we must also evaluate the conformal factor $\omega$, which
relates the metric on the $x^\alpha=(u,r,x^A)$ coordinates to the
metric on a Bondi frame at ${\cal I}^{+}$~\cite{news}. (The relevant
expressions are given in Appendix B of Ref.~\cite{news} and we will not
repeat them here.) Inspection of Eqns.~(B1)-(B6) of \cite{news}
reveals that the second angular derivatives of the conformal factor
$\omega$ enter in the calculation as well. To remove these derivatives, we
introduce the auxiliary variable ${\cal W}=\eth\omega$. Since $\omega$ is
defined solely on ${\cal I}^{+}$, we use the consistency relation ${\cal
W}_{,u}=\eth\omega_{,u}$ to propagate ${\cal W}$ along the generators
of ${\cal I}^{+}$, initializing it with ${\cal W}_{0}=\eth\omega_{0}$,
where $\omega_{0}=\omega(x^A,u=u_0)$.

\section{Numerical integration}
\label{sec:numeric}

We use the compactified coordinate $x=r/(R+r)$, such that $x=1$
at future null infinity, ${\cal I}^{+}$. Thus, the relevant radial
derivatives can be written as $\partial_r=(1-x)^2/R\partial_x$,
$r\partial_r=x(1-x)\partial_x$ and $r^2\partial_r=Rx^2\partial_x$, all
of which are regular at $x=1$. As in~\cite{cce}, we carry out the radial
integration with the right-hand side of Eqs.~(\ref{eq:Q}) evaluated at
midpoints of the radial grid,

\begin{equation}
   Q_{i} \left( x_{i} (1-x_{i}) + \Delta x \right)
  -Q_{i-1} \left( x_{i-1} (1-x_{i-1}) - \Delta x \right)
  = \frac{\left( r^2 Q \right)_{,r}}{r}|_{i-\frac{1}{2}} \Delta x,
  \quad \quad \label{eq:Qnum}
\end{equation}
and the same treatment is applied to Eq.~(\ref{eq:W}).  Note that the
hypersurface and evolution equations are manifestly regular at ${\cal
I}^{+}$ when expressed in the variable $x$, except for the terms
$-4 r B$ in Eq.~(\ref{eq:Q}) and $2r\bar\eth U$ in Eq.~(\ref{eq:W})
respectively, which have an extra factor of $r$. This factor,
which would make them singular at ${\cal I}^{+}$, is effectively
canceled by the corresponding factor of $1/r$ which appears in the
right-hand-side of Eq.~(\ref{eq:Qnum}), hence the equations are regular
for $r>0$. Note also the limiting form of Eq.~(\ref{eq:Q}) at ${\cal
I}^{+}$, $Q=-2 B$, which is useful in the discretization of $U_{,r}$.
For details on the numerical implementation of the {\it eth} operators,
see Ref.~\cite{eth}. The radial integration algorithm is explained
in detail in Refs.~\cite{axisymmetric,cce,news,disip} and we will not
cover it here. A departure from~\cite{news} is that we use a 3-step
iterative Crank-Nicholson scheme~\cite{sfrg} to ensure stability of the
time evolution~\cite{teuk}. For the time integration of the conformal
factor we use a combination of second-order Runge-Kutta and mid-point
rule~\cite{recipes} integration schemes.

\section{Waveforms from a fissioning white hole}
\label{sec:applications}

We have implemented the system of equations presented in
Sec.~\ref{sec:reduction}, and used it in the calculation of the
waveforms emitted by a fissioning white hole. The details of that
calculation are beyond the scope of this work and will appear
elsewhere~\cite{white,merger}. The spacetime exterior to a white hole
horizon is computed, by solving a double-null initial-boundary value
problem.  Boundary data is given on a white hole horizon, and initial
data is specified on an outgoing null surface emanating from the
horizon.

A procedure for the construction of the complete boundary data, {\it
i.e.} the intrinsic and extrinsic geometry of the white hole horizon,
and a description of how to obtain the boundary data necessary to compute
the exterior space-time via characteristic evolution can be found in
Ref.~\cite{horizon}. Some regularization of the equations is needed to
ensure we deal only with fields which are regular in the entire exterior
spacetime~\cite{white}.

By reducing the system of equations presented in
Refs.~\cite{horizon,white} into the form presented in
Sec.~\ref{sec:reduction}, with the addition of the auxiliary variables
$B=\eth\beta$, $\nu=\bar\eth J$ and $k=\eth K$, we are able to remove
the angular oscillation clearly present in the metric function $W$ when
this is computed with the standard approach.

We take as our benchmark case the most nonlinear (highest eccentricity)
case considered in Ref.~\cite{horizon}, with an eccentricity parameter
of $\epsilon=10^{-2}$. This corresponds to a case where the white hole
horizon pinches off before the expansion of the outward null rays goes
to zero, a necessary condition for a Bondi evolution forward in time
throughout the pre-fission period.

\begin{figure}
\centerline{\epsfysize=7.5in\epsfbox{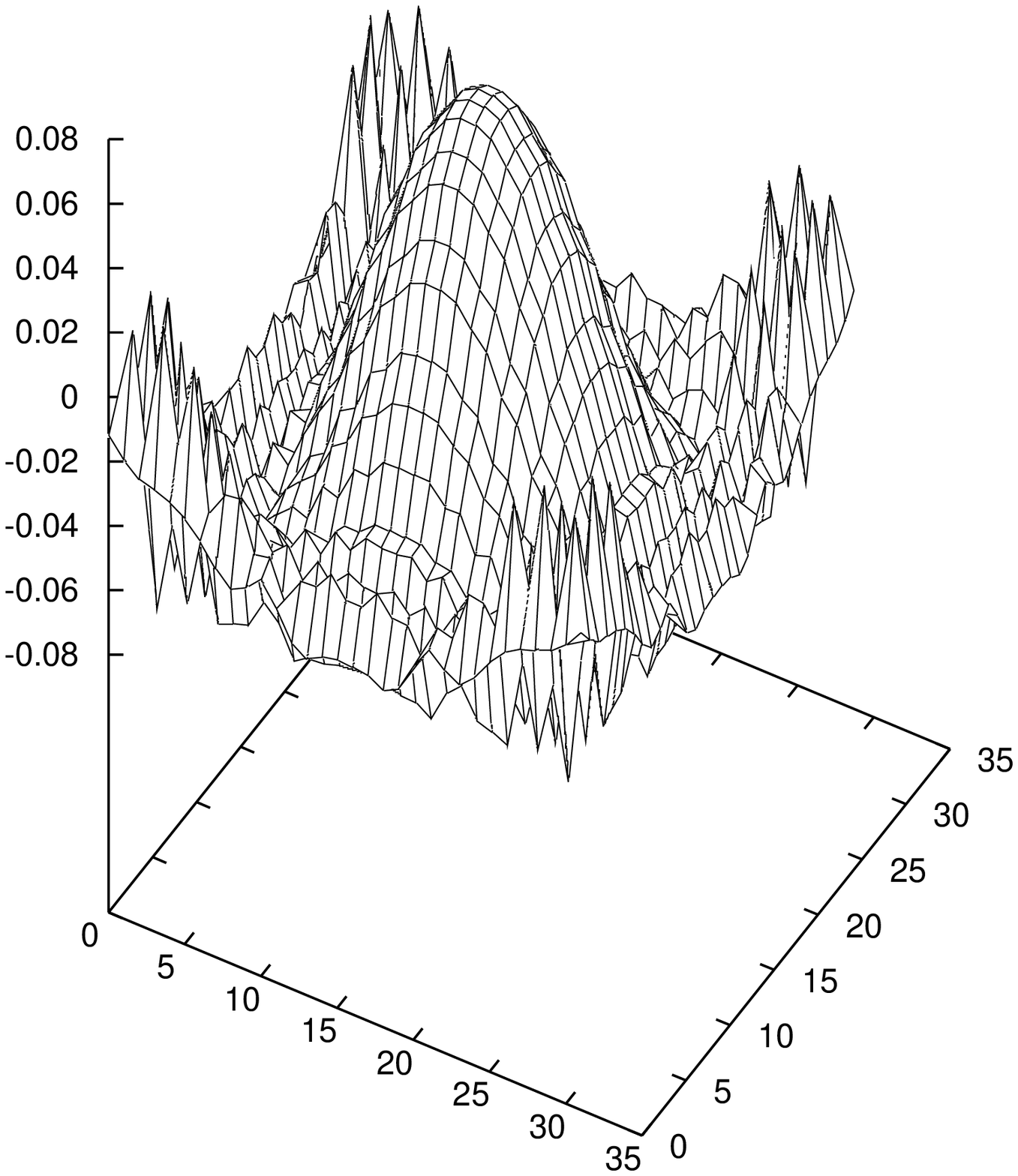}}
\caption{The metric function $W$ at null infinity, displaying the angular
oscillation mode, a consequence of the second angular derivatives in
the original system of equations. While the code is not unstable in
the usual sense, the accuracy of the computation cannot be improved by
increasing the grid resolution.}
\label{fig:Wold}
\end{figure}

\begin{figure}
\centerline{\epsfysize=7.5in\epsfbox{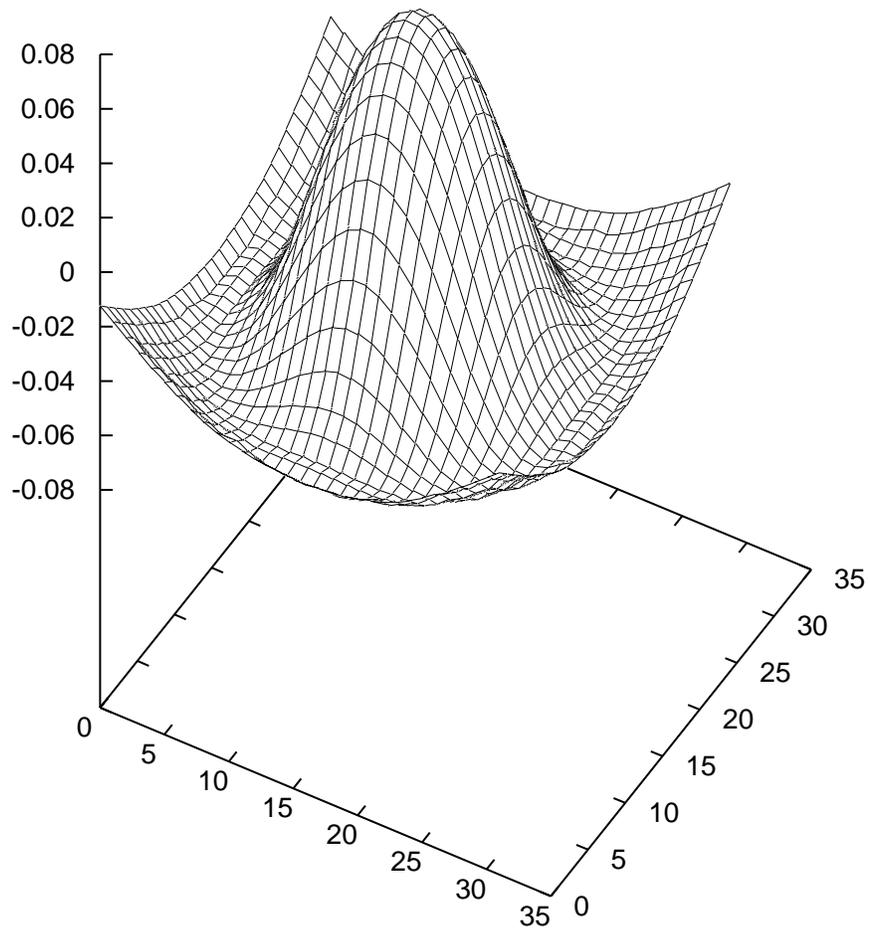}}
\caption{The variable $W$ at null infinity, when the second angular
derivatives are removed by introducing the auxiliary variables
$B=\eth\beta$, $\nu=\bar\eth J$ and $k=\eth K$. The figure shows that
the angular oscillation mode has been eliminated.}
\label{fig:Wnew}
\end{figure}

Figures~\ref{fig:Wold} and ~\ref{fig:Wnew} show the metric function $W$
at ${\cal I}^{+}$on the north stereographic patch for both approaches.
There is a clearly visible high frequency angular mode present in the
edge of the patch in the calculation performed with the standard approach,
as seen in Fig.~\ref{fig:Wold}. This oscillatory mode is clearly absent
from Fig.~\ref{fig:Wnew}.

Using both approaches (standard and reduced), we have computed the
waveforms, which correspond to the radiation emitted by a highly distorted
white hole. In the time-reversed scenario, this would correspond to
the radiation incident from ${\cal I}^{-}$, and the white-hole horizon,
to the post-merger phase of a black hole horizon formed during a binary
black hole collision. How to use this information to control the amount
of incoming radiation and to gain insight into the radiation emitted
during the post-merger phase of a black hole collision will be discussed
elsewhere~\cite{close,white,merger}.

\begin{figure}
\centerline{\epsfysize=7.5in\epsfbox{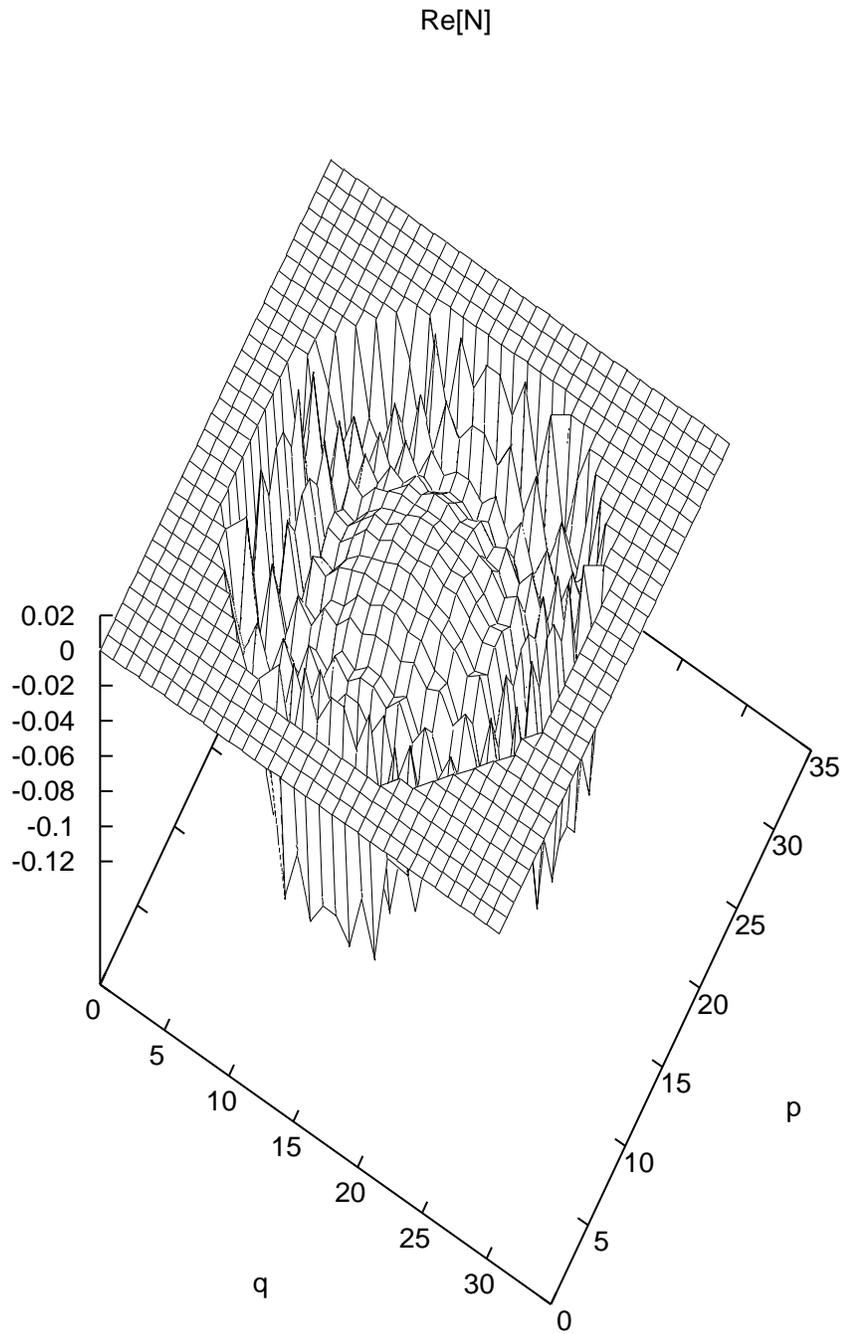}}
\caption{The real part of the Bondi news, $\Re[N]$, displays also
a pronounced angular oscillation mode. The part of the north
stereographic patch below the equator has been masked out.}
\label{fig:reNold}
\end{figure}

\begin{figure}
\centerline{\epsfysize=7.5in\epsfbox{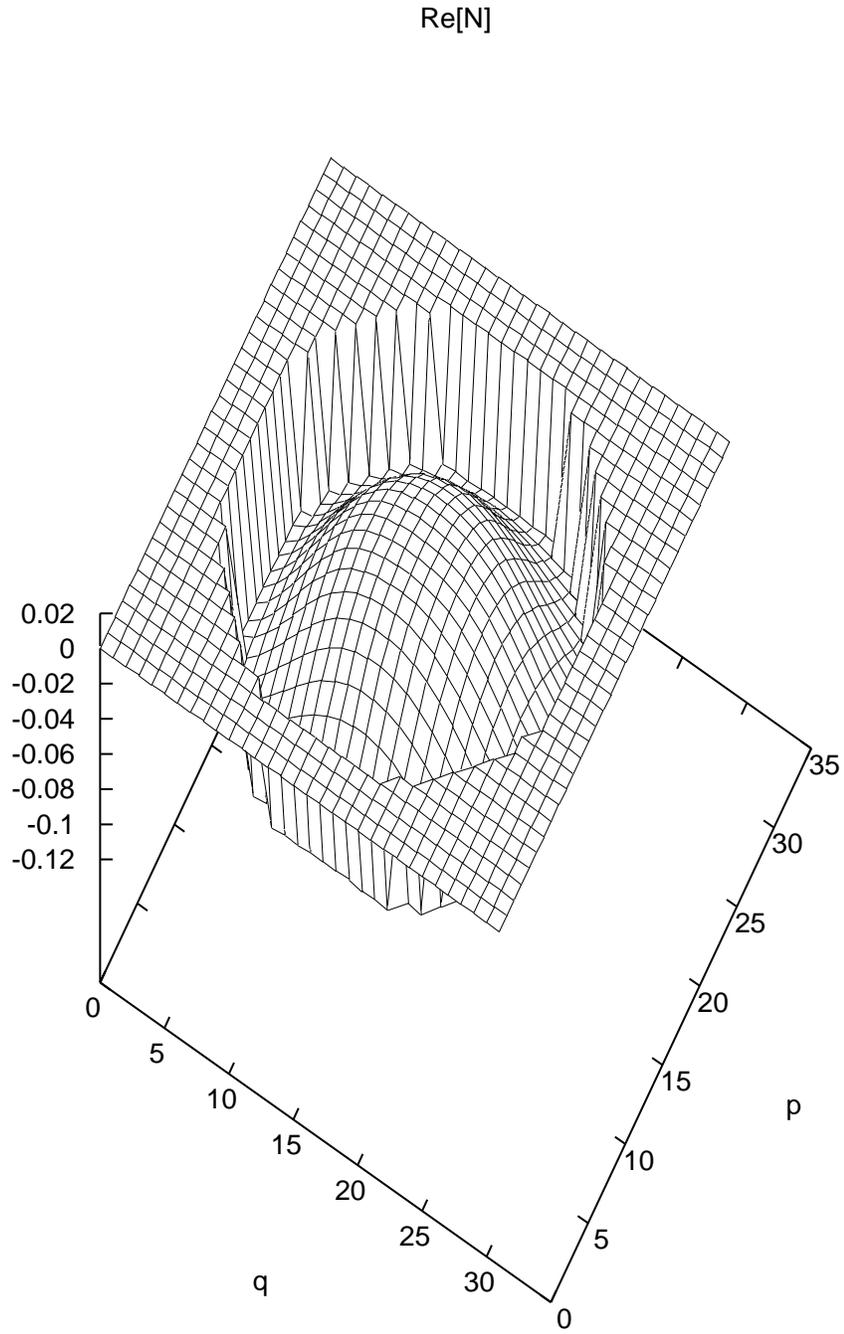}}
\caption{The real part of the Bondi news, $\Re[N]$ as computed from the
metric functions obtained from the numerical evolution of the modified
system of equations of Sec.~\ref{sec:reduction}. Comparison with
Fig.~\ref{fig:reNold} makes it clear that the angular oscillation mode is
no longer present.}
\label{fig:reNnew}
\end{figure}

Our motivation in introducing the partially reduced system was to
control the numerical difficulties encountered in the computation of the
radiation, as encoded in the Bondi news.  The quality of the numerical
waveforms is greatly improved in the partially reduced system. This is
readily apparent comparing Figs.~\ref{fig:reNold} and \ref{fig:reNnew},
which display the real part of the Bondi News, at a fixed time, for
the standard and the reduced form of the equations. (In the plots,
the region of the patch below the equator has been masked out.) While
the evolution of the exterior spacetime is stable with the standard
approach, the numerical noise swamps the signal. The noise is more
pronounced near the edges of the stereographic patch, and it propagates
to the interior. Figure~\ref{fig:reNnew} shows that, with the partially
reduced system, the Bondi News at ${\cal I}^{+}$ is perfectly smooth.

\section*{Conclusions}

We have introduced a system of equations which takes the traditional
null cone formulation and casts it as a system of equations which is of
first differential order in the angular variables.

The effectiveness of this approach is clearly demonstrated by comparing
the metric functions at null infinity, and specially the waveforms at
${\cal I}^{+}$ when computed via this approach versus those obtained
with the traditional characteristic approach, in which second-order
angular derivatives are present.

We have thus shown that the partially reduced system of equations is highly
effective in suppressing numerical errors, which otherwise would adversely
affect the calculations of waveforms in highly nonlinear scenarios.

We are currently using this approach to compute gravitational waveforms
emitted during a white hole fission~\cite{white} and during the
post-merger phase of a binary black hole collision~\cite{merger}, and
we are studying their application to the characteristic problem with
matter sources.

\acknowledgements

We thank J.~Winicour for a careful reading of the manuscript. We have
benefited from conversations with S.~Frittelli, and with N.T.~Bishop
and L.~Lehner. This work has been supported by NSF PHY 9800731 to the
University of Pittsburgh. Computer time for this project was provided
by the Department of Physics and Astronomy, by the Pittsburgh
Supercomputing Center and by NPACI.

\newpage

\newpage

\newpage

\newpage

\end{document}